\documentclass[aps,pra,floats,floatfix,twocolumn]{revtex4}

\usepackage{ifthen}
\usepackage{ifpdf}
\usepackage{color}
\ifpdf
\usepackage{graphicx}
\usepackage{epstopdf}
\else
\usepackage{graphicx}
\usepackage{epsfig}
\fi
\usepackage{latexsym}
\usepackage{amsmath}
\usepackage{amssymb}
\usepackage{bm}
\usepackage{wasysym}

\usepackage{latexsym}
\usepackage{amsmath}
\usepackage{amssymb}
\usepackage{bm}
\usepackage{wasysym}

\begin{document}
\title{Interaction-induced instability and chaos in the photoassociative stimulated Raman adiabatic passage from atomic to molecular Bose-Einstein condensates}

\author{Amit Dey$^1$ and Amichay Vardi$^2$}
\address{$^1$International Centre for Theoretical Sciences - Tata Institute of Fundamental Research, Bengaluru Ð 560089, India}
\address{$^2$Department of Chemistry, Ben-Gurion University of the Negev, Beer-Sheva 8410501, Israel}

%\pacs{
%{71.38.-k, 03.65.Yz, 05.70.Ln, 85.35.Be}}
%03.67.Pp, 75.10.Jm, 87.10.Hk}}
\date{\today}
\begin{abstract}
We study the effect of interactions on the conversion of atomic -to molecular Bose-Einstein condensates via stimulated Raman adiabatic passage. Both energetic instability during avoided crossings and dynamical instability during chaotic intervals limit adiabaticity and impose {\em low} sweep-rate boundaries on the efficiency of the process. For the diabatic traverse of avoided crossings, we find a reciprocal power-law dependence of the final unconverted population on sweep rate. For the traverse of chaos, we find a sharp low-rate boundary determined by the dynamical instability parameters. The interplay of these two mechanisms determines which instability controls the failure of molecular production. A judicious choice of sweep parameters is hence required to restore the process efficiency.
\end{abstract}
\maketitle
%\tableofcontents

{\section{Introduction}}\label{intro}

%atom-molecule experiments.
Chemical bonding was demonstrated in several quantum-gas and cold-atoms experiments \cite{Wynar00,Donley02,Regal03,Greiner03,Durr04,Hodby05,Winkler05,Syassen07,Shapiro07,Ospelkaus08}.  Leading schemes rely on magnetically controlled Feshbach resonances \cite{Donley02,Regal03,Durr04,Hodby05,Syassen07,Kohler06,Chin10} or on photoassociation via stimulated optical Raman transitions  \cite{Wynar00,Winkler05,Shapiro07}.  The possibility of hybrid Bose-Einstein condensates (BECs), described  by atom-molecule models \cite{Drummond98,Javanainen98,Javanainen99,Kostrun00,Timmermans99a,Timmermans99b,Yurovsky00,Vardi01,Anglin03,Kheruntsyan05,Meiser05,Tikhonenkov08,Mackie08,Davis08}  thus opens the way to a new form of coherent 'superchemistry' \cite{Heinzen00,Moore02,Richter15,Balakrishnan16} in which collective effects dominate reaction outcomes, as well as to molecular quantum computation \cite{Tesch02,DeMille02}.

%%%%%%%%%%%%%%%%%%%%%%%%%%%%
\begin{figure}[t]
\centering
\includegraphics[width=3.5in]{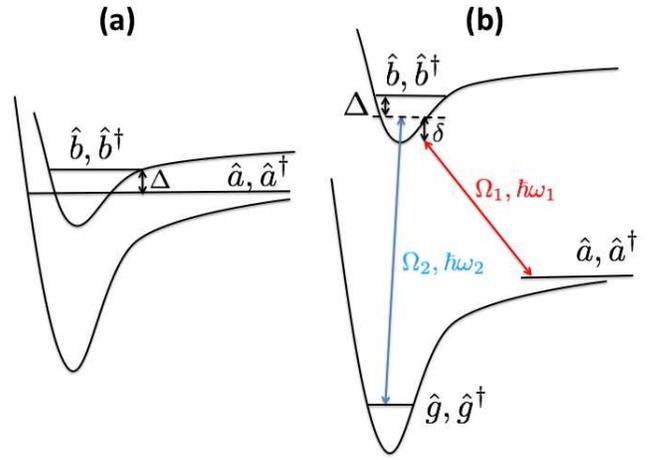} 
\caption{(color online)  Schematics of atom-molecule BEC coupling: (a) Feshbach resonance, (b) Photoassociation.}

\label{f1}
\end{figure}
%%%%%%%%%%%%%%%%%%%%%%%%%%%%

%adiabatic atom-molecule schemes
Of particular interest are schemes for adiabatic atom-molecule conversion. In the Feshbach resonance case, the paradigm is a nonlinear Landau-Zener (LZ)  sweep through resonance \cite{Ishkanyan04,Altman05,Pazy05,Tikhonenkov06,Liu08,liu08b} involving one atomic and one molecular mode (Fig.~\ref{f1}a). In comparison, stimulated Raman photoassociation couples three second quantized modes. Atoms are coupled to ground state molecules via an intermediate excited molecular state (Fig.~\ref{f1}b). Adiabatic Photoassociation \cite{Shapiro07} can thus be induced via either rapid adiabatic passage (RAP), i.e. by sweeping the frequencies of the coupling lasers through two photon resonance, or by pulsing their intensity in a counter-intuitive Stokes-before-pump order while maintaining two photon resonance, a technique known as {\em Stimulated Raman Adiabatic Passage} (STIRAP)\cite{Gaubatz90,Vitanov17,Vardi97,Julienne98,Mackie00,Hope01,Drummond02,Mackie04,Mackie05,Drummond05,Pu07,Meng08}. Since photoassociative STIRAP proceeds via the adiabatic manipulation of an atom-molecule dark state \cite{Winkler05} that does not project onto the intermediate excited molecular state, its advantage over RAP is the avoidance of spontaneous emission losses.

%Possibility of Chaos
One aspect of atom-molecule dynamics that has been neglected so far, is the emergence of {\em dynamical chaos}. The two-mode Feshbach scenario involves just one classical degree of freedom, hence its integrability is maintained throughout the sweep process. By contrast, photoassociative adiabatic passage requires two classical degrees of freedom and thus enables chaotic dynamics. 

Recently, we have studied the emergence of chaos and instability due to interactions in the standard (linear coupling) STIRAP scheme \cite{Dey18,Dey19}. We have found that the process efficiency is controlled by chaotic intervals of {\em dynamical} instability, rather than by the previously known energetic instability during nonlinear avoided crossing \cite{Graefe06}. These stochastic intervals impose {\em low} sweep-rate boundaries on the process' efficiency: In addition for the standard requirement that the control parameters be changed slowly with respect to internal characteristic frequencies, they must be varied sufficiently {\em quickly}  so as to successfully traverse the chaotic intervals before stochastic spreading takes place. 

In this paper, we revisit the process of atom-molecule STIRAP in the presence of inter-species and intra-species nonlinear interaction. While the detrimental effect of nonlinear scattering terms have been numerically investigated \cite{Hope01,Drummond02,Mackie04,Mackie05,Drummond05}, the role of the ensuing instabilities and in particular the emergence of non-integrable chaotic dynamics, have not been previously highlighted. Carrying out the stability analysis around the followed atom-molecule dark state, we find, similarly to Refs.~\cite{Dey18,Dey19}, both energetic instabilities that show up as avoided crossings in the bifurcation diagram, and intervals of chaotic dynamical instability that have no trace in the bifurcation diagram. Unlike Refs.~\cite{Dey18,Dey19}, where the process efficiency was overwhelmingly limited by dynamical instabilities, we find that depending on pulse parameters, both avoided crossings and chaos can play a role in setting the low sweep rate boundaries. In particular, the former lead to a reciprocal power-law dependence of the remnant population on the sweep rate, whereas the latter introduces a sharp low rate boundary that can be determined from the Bogoliubov stability parameters.

The paper is arranged as follows. In Section~II we present the atom-molecule STIRAP model. In Section~III we work out the adiabatic stationary point (SP) solutions and in Section~IV present numerical simulations of slow but finite sweep-rate dynamics, demonstrating the low sweep-rate boundaries on efficient molecular BEC production. Energetic and dynamical instabilities are determined in Section~V using the Bogoliubov perturbative formalism. Poincar\'{e} sections (PCS) are used in section~VI to associate intervals of dynamical instability with chaotic quasi-static dynamics. Using this information, the numerical results and their dependence on the sweep rate, are explained in Section~\ref{rate}. Conclusions are presented in section \ref{conclusion}. 

{\section{Photoassociative STIRAP}}\label{model}

We consider the photoassociation scheme schematically illustrated in Fig.~\ref{f1}b. An atomic BEC is converted into a ground state molecular BEC, via a STIRAP sequence in a $\Lambda$ configuration involving an excited molecular mode. The STIRAP sequence proceeds via the usual counter-intuitive Stokes-before-pump pulse sequence \cite{Vitanov17} while maintaining two-photon resonance, thus guiding an atom-molecule dark state from mode 'a' to mode 'g' without significant projection onto mode 'b' throughout the process. Under the assumption that photoassociation is fast with respect to motional timescales, the pertinent Hamiltonian is given by the three-mode parametric Bose-Hubbard model \cite{Mackie00,Hope01},
\begin{eqnarray}
 \hat{H}&=& \Delta \hat{b}^{\dagger}\hat{b} -\frac{\Omega_1(t)}{2\sqrt{N}} (\hat{a}\hat{a}\hat{b}^{\dagger}+h.c.)-\frac{\Omega_2(t)}{2}(\hat{b}^{\dagger}\hat{g}+h.c.) \nonumber \\
 &&~+\frac{1}{2}(U_{aa}\hat{n}^2_a+U_{bb}\hat{n}^2_b+U_{gg}\hat{n}^2_g)+U_{ag}\hat{n}_{a}\hat{n}_g,
 \label{model}
\end{eqnarray}
where $\hat{a}$, $\hat{b}$ and $\hat{g}$ are bosonic annihilation operators for particles in the corresponding modes, $\hat{n}_\alpha=\hat{\alpha}^{\dagger}\hat{\alpha}$ are mode populations, $\Omega_{1,2}$ are the 'pump' atom-molecule (a-b) and 'Stokes' molecule-molecule (b-g) couplings, respectively, and $\Delta$ is the one-photon detuning. 
The intraspecies interaction strengths are denoted by $U_{aa}$, $U_{bb}$, and $U_{gg}$, while the atom-molecule scattering strength is $U_{ag}$. Atom-molecule interaction involving excited molecules is neglected, as the intermediate mode's population remains small throughout the photoassociation process. To implement STIRAP, the couplings are taken to be Gaussian pulses $\Omega_{1,2}(t)=K {\rm exp \Big[-(x-x_{1,2})^2\Big]}$ in a 'counter-intuitive'  sequence ($x_{1}>x_2$) with $x=t/\tau$ where $\tau$ is the pulse width. Throughout the manuscript peak times are $x_1=4$ and $x_2=2.33$.

In the classical mean-field limit, obtained as $N\rightarrow\infty$ at fixed $U_{\alpha\beta}$N, the field operators are replaced by $c$-numbers. Rescaling the amplitudes as $a\rightarrow a/\sqrt{N}$, $b\rightarrow b/\sqrt{N}$, $g\rightarrow g/\sqrt{N}$,  and time as $t\rightarrow Kt$, the classical Hamiltonian reads,
\begin{eqnarray}
 H_x&=&\delta |b|^2-\frac{1}{2}[\bar{\Omega}_1(x)(a^{*2}b+a^2 b^*) + \bar{\Omega}_2(x) (b^*g+g^* b)  ]\nonumber \\
 &&~+\frac{1}{2} [ u_{aa}|a|^4 + u_{bb} |b|^4 + u_{gg} |g|^4  ] + u_{ag} |a|^2 |g|^2,
 \label{H_class}
\end{eqnarray}
where the dimensionless parameters are $\delta=\Delta/K$, $\bar{\Omega}_{1,2}(x)=\Omega_{1,2}(x)/K$ and $u_{\alpha\beta}=U_{\alpha\beta}N/K$. Below we employ interaction parameter ratios that are roughly consistent with those of Rubidium ($^{87}$Rb), such that $u_{bb}=u_{gg}=u_{aa}/2$, and $u_{ag}=1.25u_{aa}$ \cite{Drummond02,Mackie04}. The atom-atom interaction is taken to be $u_{aa}=0.2c$, where c corresponds to the relative the strength of interaction with respect to the linear laser coupling between modes. An additional motional constant is provided by total number conservation, i.e. $|a|^2+2|b|^2+2|g|^2=1$.\\

\section{Stationary points and bifurcation diagrams}\label{bifurcation}
The adiabatic stationary points (SPs) ${\bf s}^T=(a_s,b_s,g_s)$, of the classical dynamics at any fixed value of the control parameter $x$, are obtained by extremizing $H_x-(\mu_x/K)(|a|^2+2|b|^2+2|g|^2 )$, where $\mu_x$ denotes the chemical potential at $x$, thus obtaining the time-independent equations,
\begin{equation}
[\mathcal{H}_{1}({\bf s})+\mathcal{H}_{2}({\bf s})]{\bf s}={\boldsymbol \mu} {\bf s},
\end{equation}
where
\begin{eqnarray}
%i\dot{a}&=&-\frac{\Omega_{1}}{K}a^* b+u_{aa}|a|^2a+u_{ag}|g|^2 a, \\
%i\dot{b}&=&\Delta b-\frac{\Omega_2}{2K}g-\frac{\Omega_1}{2K}a^2+u_{bb}|b|^2b,\\
%i\dot{g}&=&-\frac{\Omega_2}{2K}b+u_{gg}|g|^2g+u_{ag}|a|^2g,
 \mathcal{H}_1({\bf s})=  
\begin{pmatrix}
0 & -\bar{\Omega}_1 a^* & 0 \\
-\frac{{\bar{\Omega}_1}}{2} a & \delta & -\frac{\bar{\Omega}_2}{2} \\
0 & -\frac{\bar{\Omega}_2}{2} & 0
\end{pmatrix},
\label{H1}
\end{eqnarray}
\begin{eqnarray}
\mathcal{H}_2({\bf s})=  
\begin{pmatrix}
u_{aa}|a|^2+u_{ag}|g|^2 & 0 & 0 \\
0& u_{bb}|b|^2 & 0 \\
0 & 0 & u_{gg}|g|^2+ u_{ag}|a|^2
\end{pmatrix},
\end{eqnarray}
and
\begin{eqnarray}
{\boldsymbol \mu}=  
\begin{pmatrix}
 \mu/K & 0 & 0 \\
0& 2\mu/K & 0 \\
0 & 0 & 2\mu/K
\end{pmatrix}.
\end{eqnarray}
The energy of each stationary solution is then evaluated as $E_{SP}(x)=H_x({\bf s}(x))$. It should be noted that contrary to linear STIRAP, one typically obtains more than three SPs, up to a maximum of eight. 

\begin{figure}[t]
  \centering
   \includegraphics[width=3.2in]{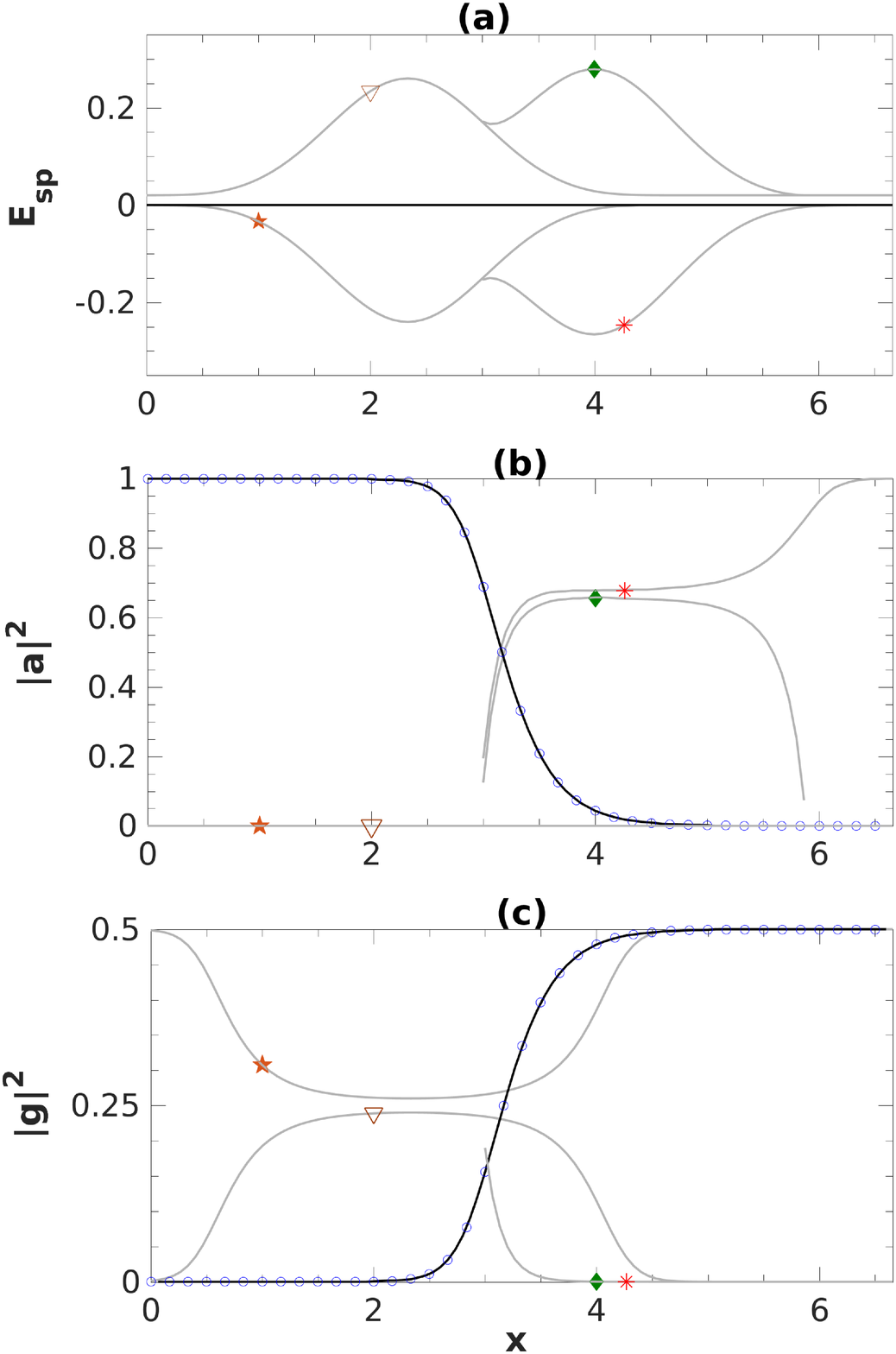}
  \caption{Atom-molecule STIRAP dynamics in the absence of interaction ($c=0$): (a) Stationary point energies; (b) Stationary point atomic population; (c) Stationary point target molecular population. Markers distinguish between different SPs and associate the projections with a given adiabatic energy curve. The followed (dark-state) SP is marked by a solid thick black line. Circles mark the observed successful conversion dynamics at a rate of $\dot{x}=6.6\times10^{-5}$. There is no deterioration of conversion efficiency at slower sweep rates.}
  \label{f2}
\end{figure}

\begin{figure}[t]
  \centering
   \includegraphics[width=3in]{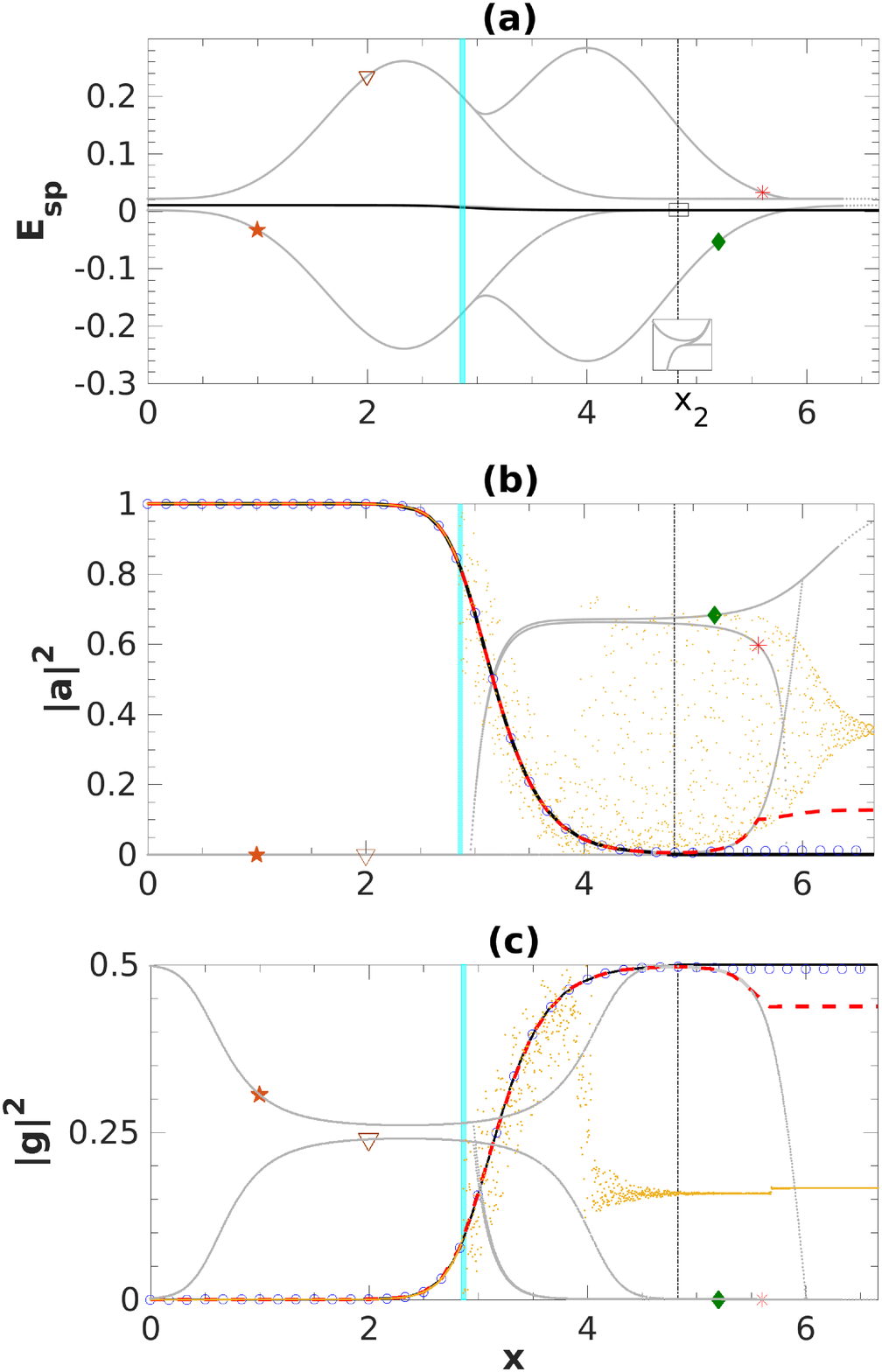}
  \caption{Atom-molecule STIRAP dynamics for $c=0.1$. Same convention as in Fig.~\ref{f2} is used to present the SP solutions. However, the dynamics at a rate of $\dot{x}=6.6\times10^{-4}$ (blue circles), $\dot{x}=6.6\times10^{-6}$ (dashed red line), and $\dot{x}=3.3\times10^{-7}$ (yellow dotted line), demonstrate that interaction leads to the failure of atom-to-molecule conversion at {\em slow} sweep rates. The failure onset points at different slow sweep rates match the chaotic interval (cyan shaded band) and the $x_2$ avoided crossing (vertical dash-dotted line), determined from the stability analysis in section \ref{stability}.}
  \label{f3}
\end{figure}

\begin{figure}[h]
  \centering
  \includegraphics[width=3.0in]{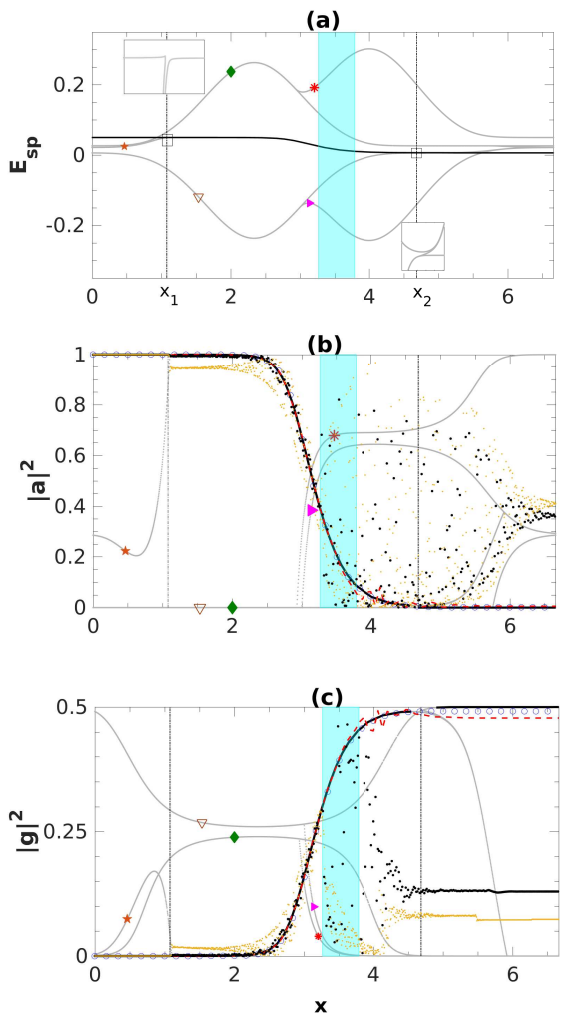}
  \caption{Atom-molecule STIRAP dynamics for $c=0.5$. SP solutions are shown using the same convention as in Fig.~\ref{f2}. Dynamics at rates of $\dot{x}=3.3\times10^{-2}$ (blue circles) gives near 100\% conversion efficiency.  Slower sweep rates of $\dot{x}=3.3\times10^{-3}$ (dashed red line), $\dot{x}=3.3\times10^{-5}$ (bold black dots) and $\dot{x}=3.3\times10^{-7}$ (light yelow dots) show failure at the chaotic interval (shaded cyan band) or at either one of the avoided crossings at $x_{1,2}$ (vertical dash-dotted lines).}
  \label{f4}
\end{figure}

Plotting $E_{SP}$ for all stationary solutions as a function of $x$, we obtain the {\em bifurcation diagrams} shown in the top panels of Figs.~\ref{f2}-\ref{f4} for various strengths of the interaction. The corresponding $a_s$ and $g_s$ populations of the stationary solutions are presented in the lower panels of the same figures. 
 
{\section{Slow-sweep dynamics}}\label{dynamics}
Starting with the non-interacting case (Fig.~\ref{f2}) one still obtains a proliferation of SPs due to the nonlinear atom-molecule coupling. However, one adiabatic solution, marked by bold lines in Fig.~\ref{f2}, is an atom-molecule dark state, transforming from an all-atomic state $|a|^2=1$ at $x=0$ to an all-molecular state $|g|^2=1$ at $x\rightarrow \infty$, while maintaining $|b|^2=0$ for all $x$. Plotting the population dynamics, during a numerical simulation with a finite-time sweep at $\dot{x}=6.6\times10^{-5}$ (circle markers in Fig.~\ref{f2}a), the system successfully follows the adiabatic ground state, eventually producing a molecular condensate with near-unit efficiency. This result is representative for all sweep rates below the standard adiabatic threshold $K\tau\gg 1$, hence there is no low rate boundary on the photoassociative STIRAP efficiency.

The situation is quite different in the presence of even a relatively moderate ($c=0.1$) interaction (see Fig.~\ref{f3}). While the atom-molecule dark state is still present, we observe a surprising {\em slow sweep} deterioration of the transfer efficiency, similar to the results of \cite{Dey18,Dey19}.  The fastest sweep shown ($\dot{x}=6.6\times10^{-4}$, blue circles) guarantees successful conversion. However, at a {\em slower} rate ($\dot{x}=6.6\times10^{-6}$, dashed red line) the population dynamics deviates from the followed adiabatic state towards the end of the process. At an even slower rate ($\dot{x}=3.3\times10^{-7}$, orange dotted line) the breakdown point is shifted to a much earlier stage.

the dynamics in the presence of stronger interaction ($c=0.5$) is presented in Fig.~\ref{f4}. Here, there are {\em three} distinct breakdown points obtained at different sweep rates. the fastest sweep has close to 100\% efficiency, but slower sweeps fail at either of the marked vertical lines, or at the shaded interval between them. Below we show that the vertical lines in Fig.~\ref{f3} and Fig.~\ref{f4} correspond to avoided crossings in the bifurcation diagram, indicating an {\em energetic instability}, while the shaded regions mark intervals of {\em dynamical instability} due to chaotic dynamics around the SP.  Since each such instability could be traversed using a sufficiently {\em fast} sweep, and since the threshold rates differ for different instabilities, we obtain that different instabilities manifest at different sweep rates.

We note that for larger values of $c$, the interaction can not be treated as a perturbation of the photoassociative STIRAP dynamics and an atom-molecule dark state is no longer available, altogether precluding the possibility of efficient atom-molecule transfer, regardless of pulse duration.\\

\section{Stability analysis}\label{stability}

In order to characterize the effect of interactions on the stability of the followed SP during photoassociative STIRAP, we carry out Bogoliubov stability analysis \cite{bogo,pethick} around it. Perturbing the state vector about the SP solution, so that ${\bf s}(t)\Rightarrow {\bf s}^{}_{SP}+\delta {\bf s}(t)$, transforming $\delta {\bf s}(t) \equiv {\bf u} e^{-i\omega t} - {\bf v}^* e^{i\omega t}$ and linearizing the resulting dynamical equations, we obtain the Bogoliubov equations for the quasiparticle modes $\{ {\bf u}, {\bf v} \}$ and their characteristic frequencies $\omega$,

\begin{eqnarray}
 &&\Big[{\boldsymbol \mu}+\omega-\mathcal{H}({\bf s}^{}_{SP})+\mathcal{A}({\bf s}^{}_{SP})\Big]{\bf u} \nonumber \\ 
 &&~~~~~~~~~~~~~~~~~~+ \Big[\mathcal{H}_3({\bf s}^{}_{SP}) -\mathcal{B}({\bf s}^{}_{SP})\Big]{\bf v}=0, 
 \\
 &&\Big[{\boldsymbol \mu}-\omega-\mathcal{H}({\bf s}^{}_{SP})+\mathcal{A}({\bf s}^{}_{SP})\Big]{\bf v} \nonumber \\ 
 &&~~~~~~~~~~~~~~~~~~+ \Big[\mathcal{H}_3({\bf s}^{}_{SP}) -\mathcal{B}({\bf s}^{}_{SP})\Big]{\bf u}=0, 
\end{eqnarray}
where $\mathcal{H}=\mathcal{H}_{1} +\mathcal{H}_2+\mathcal{H}_3$ with the definitions

\begin{eqnarray}
 \mathcal{H}_3({\bf s}^{}_{SP})&=&  
\begin{pmatrix}
u_{aa} a_{SP}^2 & 0 & u_{ag}a^{}_{SP}g^{}_{SP} \\
0& u_{bb}b_{SP}^2 & 0 \\
u_{ag}a^{}_{SP}g^{}_{SP} & 0 & u_{gg}g_{SP}^2
\end{pmatrix},\\
\mathcal{A}({\bf s}^{}_{SP})&=&  
\begin{pmatrix}
0 & 0 & 0 \\
\frac{\bar{\Omega}_1}{2}a^{}_{SP}& 0 & 0 \\
0 & 0 & 0
\end{pmatrix},\\
\mathcal{B}({\bf s}^{}_{SP})&=&
\begin{pmatrix}
\bar{\Omega}_1b^{}_{SP} & 0 & 0 \\
0& 0 & 0 \\
0 & 0 & 0
\end{pmatrix}.
\end{eqnarray}

\begin{figure}[t]
  \centering
   \includegraphics[width=3.3in]{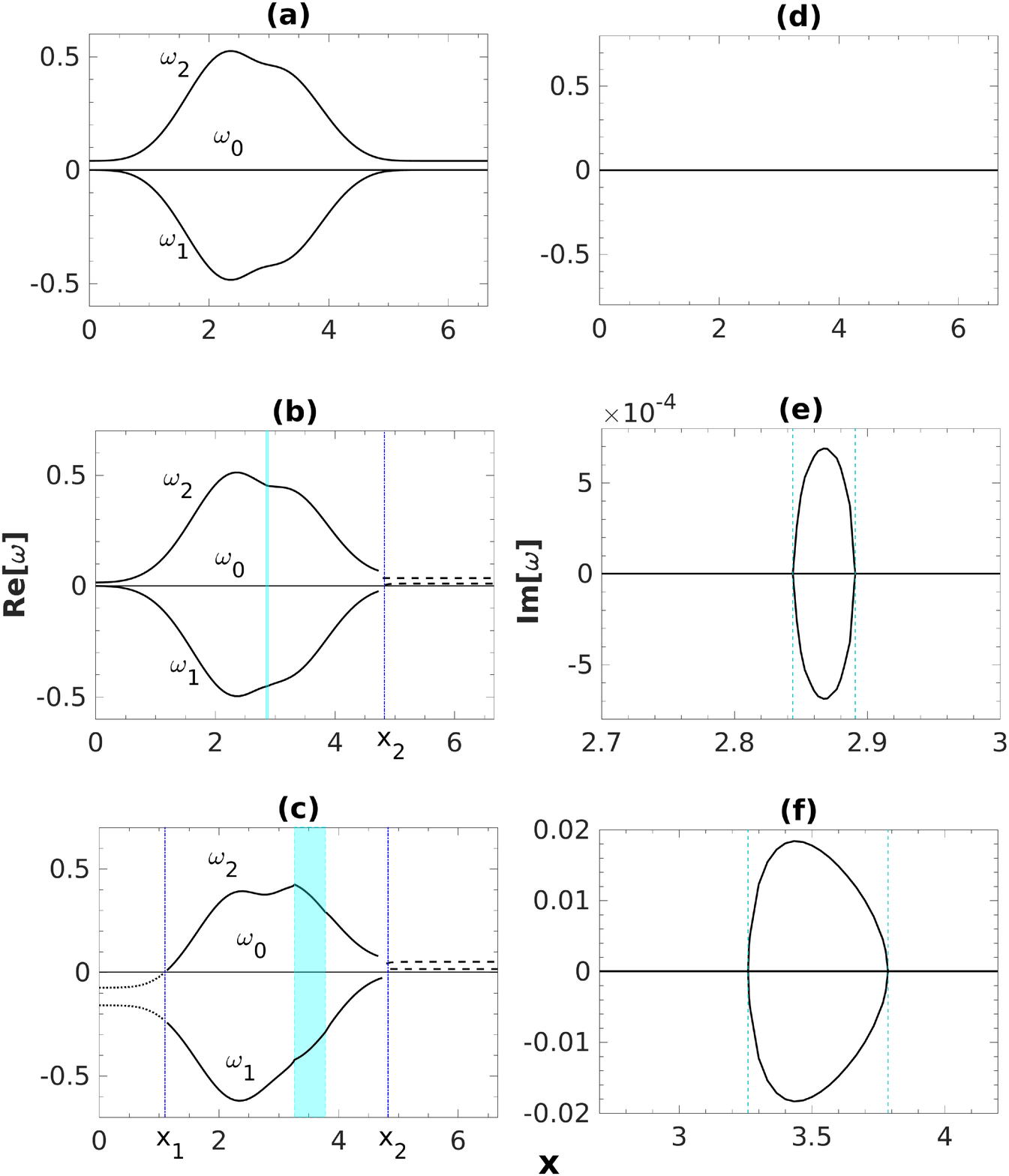}
  \caption{Characteristic Bogoliubov frequencies $\{\omega_q\}_{q=0,1,2}$ around the followed SP. The real and imaginary parts of the frequencies are plotted respectively for $c=0$ (a,d), $c=0.1$ (b,e), and $c=0.5$ (c,f). Note the $x$-axis in e,f is chosen to zoom on the region of dynamical instability, where ${\rm Im}(\omega)\neq 0$. The corresponding dynamical instability intervals are marked by shaded cyan regions in panels b,c. The dotted (dashed) parts of ${\rm Re}[\omega]$ mark the $x$ range where the followed SP is a maximum (minimum) of the energy landscape, while solid lines mark the region where the SP is an energetic saddle point. Energetic stability is lost (regained) at the avoided crossing point $x_1$ ($x_2$), marked by vertical blue lines in panels b,c.}
  \label{f5}
\end{figure}

The photoassociation atom-molecule system has three classical degrees of freedom (e.g. the amplitudes $\sqrt{n_{a,b,g}}$ and phases $\varphi_{a,b,g}$ serving as canonical action-angle coordinates). Total number conservation thus implies there would be one zero Bogoliubov mode and two non-trivial modes. When the frequencies are real, their sign determines the energetic stability of the SP. By contrast, an imaginary Bogoliubov frequency indicates the dynamical instability of the motion within the SP's energy surface. While energetic minima and maxima are always dynamically stable, energetic saddle points can be either dynamically elliptical or dynamically hyperbolic.
 
The real and imaginary parts of the resulting Bogoliubov frequencies for the interaction strengths of Figs.~\ref{f2}-\ref{f4}, are plotted as a function of $x$ in Fig.~\ref{f5}. For the noninteracting case ($c=0$, see Fig.~\ref{f5}a,d)  the followed SP remains a dynamically stable energetic saddle point ($\omega_1<0$, $\omega_2>0$, ${\rm Im}[\omega_{1,2}]=0$) throughout the process. By contrast, for $c=0.1$ (Fig.~\ref{f5}b,e) there is a transition from an energy saddle ($\omega_1<0,\omega_2>0$) to an energy minimum ($\omega_{1,2}>0$) precisely at the point $x=x_2$ where we have observed the late breakdown for the ${\dot x}=6.6\times10^{-6}$ sweep in Fig.~\ref{f3}. This transition shows up as an avoided crossing in the bifurcation diagram. Moreover, a brief interval of dynamical instability (${\rm Im}[\omega_{1,2}]\neq 0)$ precisely matches the onset of breakdown at the slowest sweep rate ${\dot x}=3.3\times10^{-7}$ in the same figure. This interval has no trace in the bifurcation diagram that only reflects the structure of the energy landscape rather than the dynamics within each energy surface.

Similarly, for the strongest interaction $c=0.5$ case (Fig.~\ref{f5}c,f), we can correlate the observed breakdown of photoassociative STIRAP in Fig.~\ref{f4} at different sweep rates, with various energetic and dynamical instabilities. Here, the followed SP starts out as an energy maximum ($\omega_{1,2}<0$), becomes a saddle ($\omega_1<0,\omega_2>0$) at $x=x_1$  and then a minimum ($\omega_{1,2}>0$) at $x=x_2$. The avoided crossings at the $x_{1,2}$ transition points explain the observed breakdown at the ${\dot x}=3.3\times10^{-7}$ and ${\dot x}=3.3\times10^{-3}$ sweeps of Fig.~\ref{f4}, respectively. Additionally, the breakdown at the intermediate rate of ${\dot x}=3.3\times10^{-5}$
in Fig.~\ref{f4} corresponds to a broad interval of strong dynamical instability.

\begin{figure}[t]
  \centering
   \includegraphics[width=3.2in]{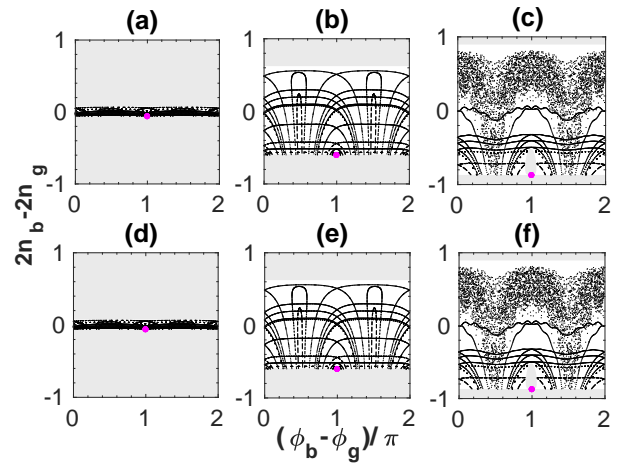}
  \caption{Poincar\'{e} sections in the $\{n_a[SP],E_{SP}\}$ plane for the noninteracting system at (a), (d) $x=2.6667$, (b), (e) $x=3.2667$, (c), (f) $x=3.6667$. In the upper row, the magenta dots correspond to the evolution of a small cloud of classical points prepared around the all-atoms SP at $x=0$, under time-dependent Hamiltonian untill the stated $x$ values and then under the frozen Hamiltonian $H(x)$ for a long time. The second row shows the quasi-static evolution of the semiclassical cloud prepared at SP$(x)$ under the fixed Hamiltonian  $H(x)$. Thus, the top row corresponds to the accumulated effects of instabilities up to the stated $x$ while the bottom row highlights the instantaneous stability/instability at the stated value of $x$. For the noninteracting system, there are no instabilities and hence no distinguishable difference between the rows.}
  \label{f6}
\end{figure}
  
\section{Poincar\'{e} sections} \label{poincare}

In order to explore the origin of the dynamical instabilities in Fig.~\ref{f5}e,f, we study the structure of the pertinent energy surfaces containing the followed SP, using Poincare cross-sections. Eliminating one degree of freedom due to number conservation, the phase-space of the atom-molecule system is four-dimensional (e.g. spanned by two population imbalances and two relative phases between modes). Fixed energy surfaces $E=E_{\rm SP}$ are thus 3D. For the finite interaction $c\neq0$ cases, we choose the dynamical coordinates to be $n_b$, $n_a-2n_g$, and $2\phi_a-\phi_g$. Poincar\'{e} sections are obtained by plotting a point in the  $\{n_a-2n_g,2\phi_a-\phi_g\}$ plane each time a trajectory hits the followed SP value of $n_b$, i.e. $n_b(x)=|b^{}_{SP}(x)|^2$. For the noninteracting ($c=0$) case, we have $n_b[{\bf SP}]=0$ for all $x$ and our conjugate variables are $2n_b-2n_g$ and $\phi_b-\phi_g$ for the surface \{$E_{SP},n_a[SP]$\}. 

\begin{figure}[t]
  \centering
   \includegraphics[width=3.2in]{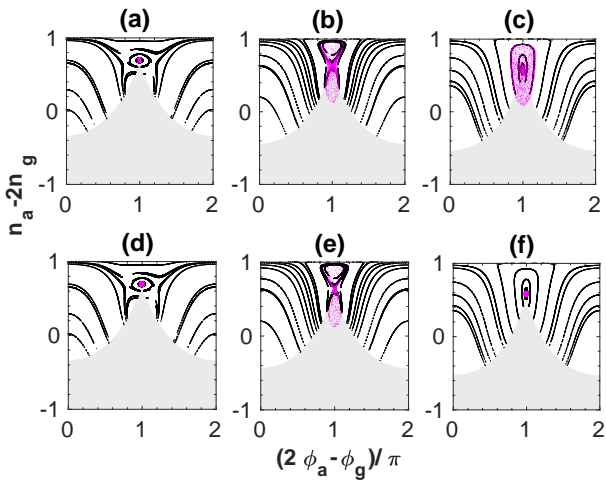}
  \caption{Poincar\'{e}  sections for $c=0.1$ in the $\{n_b[SP],E_{SP}\}$ plane, taken at :  (a), (d) $x=2.8333$, (b), (e) $x=2.8667$, (c), (f) $x=2.9$. Top (bottom) row is accumulated (instantaneous) evolution, as in Fig.~\ref{f6}. A brief chaotic interval where the semiclassical cloud stochastically spreads along a chaotic strip (panels b,e), matches the $x$ values for which dynamical instability is seen in Fig.~\ref{f5}e. While the instantaneous dynamics after this interval returns to stability (panel f), the spreading that took place within it can not be reversed (panel c) .}
  \label{f7}
\end{figure}

In Fig.~\ref{f6} we plot representative Poincare sections throughout photoassociative STIRAP in the absence of interaction. Over these sections, we overlay the semiclassical propagation of an initially localized Gaussian cloud of classical points. In the top row, we plot the free evolution up to the specified value of $x$, under the slowly varying Hamiltonian, thus showing the accumulated spreading during the propagation. In contrast, the bottom row shows the propagation under the frozen instantaneous Hamiltonian at the specified $x$, thus highlighting the precise values of $x$ where the followed SP becomes unstable.   Interestingly the energy shell of the followed SP contains chaotic regions even without interaction, due to the nonlinearity of the $a-b$ atom-molecule coupling. However, consistently with its stability analysis, the followed SP avoids these stochastic regions throughout its evolution and consequently remains localized. Since no instability is encountered, there is no discernible difference between the varying-$x$ and fixed-$x$ dynamics.

\begin{figure}[t]
  \centering
   \includegraphics[width=3.2in]{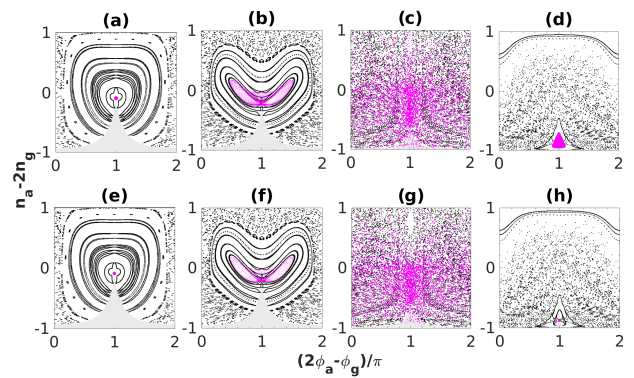}
  \caption{Poincar\'{e}  sections for $c=0.5$ in the $\{n_b[SP],E_{SP}\}$ plane, taken at : (a), (e) $x=3.2$, (b), (f) $x=3.2667$, (c), (g) $x=3.45$, (d), (h) $x=3.9$. Upper (lower) row is accumulated (instantaneous) evolution, as in Fig.~\ref{f6} and Fig~~\ref{f7}. The dynamical instability in the Bogoliubov analysis (See Fig.~\ref{f5}f) precisely matches the range of $x$ where the followed SP is embedded in the chaotic sea.}
  \label{f8}
\end{figure}

The Poincare sections in the presence of $c=0.1$ interactions are presented in Fig.~\ref{f7}. The dynamical instability at $x=2.8667$ (see Fig.~\ref{f5}) corresponds to the spreading of the semiclassical cloud over a narrow chaotic strip (Fig.~\ref{f7}b), resulting in the observed breakdown of the slowest sweep dynamics (orange pointed line)  in Fig.~\ref{f3}. Comparing panels (c) and (f), we see that in agreement with Fig.~\ref{f5}e, the SP at $x=2.9$ returns to dynamical stability, but the accumulated effect of spreading during the preceding chaotic interval can not be undone. More pronounced stochastic spreading is observed during the broader chaotic interval for $c=0.5$, as shown in Fig.~\ref{f8}.

\section{Dependence of conversion efficiency on sweep rate} \label{rate}

As noted in Section~\ref{dynamics}, the outcome of photoassociative STIRAP depends strongly on the sweep rate, with unexpected deterioration of the process efficiency for longer pulse durations. This dependence now becomes clear, since at the limit of an infinitely slow sweep the system is not able to traverse the avoided crossings at $x_1$ and $x_2$ and is also subject to stochastic motion during the chaotic interval between them. In order to recover the photoassociation efficiency, the process has to be carried out sufficiently {\em quickly} so that the avoided crossings are traversed {\em diabatically} and the chaotic interval is crossed before stochastic spreading takes place.  

\begin{figure}[t]
  \centering
   \includegraphics[width=3.5in]{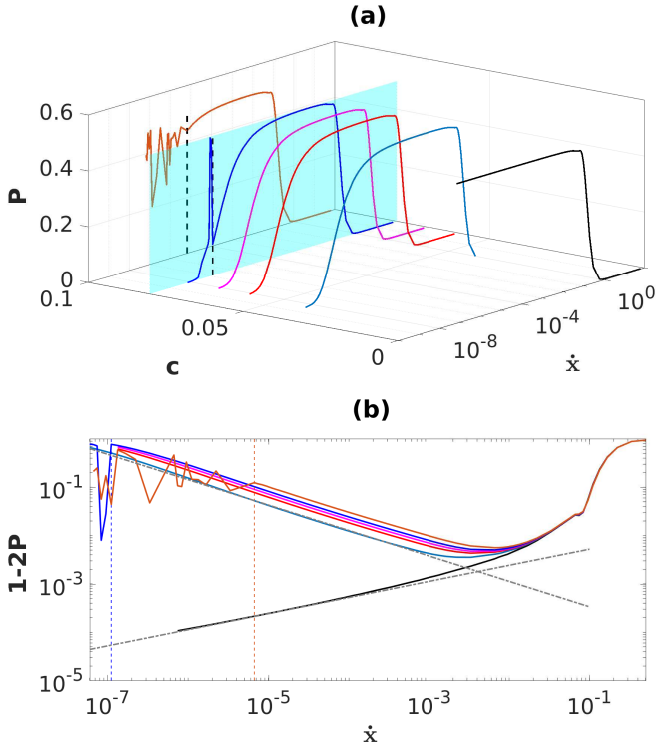}
  \caption{Sweep rate dependence of the conversion efficiency $P\equiv n_g(\infty)$ for $c=0,.04,.06,.07,.08$ and $0.1$ (self-evident in panel a, bottom to top lines in panel b), describing the transition from $x_2$-dominated breakdown to chaos-induced failure: (a) Efficieny plot in a $\dot{x}-c-P$ layout, with the cyan plane marking the interaction strength $c_{\rm chaos}$ at which chaotic intervals emerge, (b) log-log plot of the remnant population $1-2P$ at the same sweep rates, with dash-dotted lines marking $1-2P\propto{\dot x}^{1/3}$ and $1-2P\propto{\dot x}^{-1/2}$ power law dependence. The non-interacting system (black solid line) exhibits an ${\dot x}^{1/3}$ power law decrease of the remnant population as the sweep rate decreases, hence no low sweep-rate boundary. For $0<c<c_{\rm chaos}$ the power law is inverted to ${\dot x}^{-1/2}$ due to the need to quickly traverse the avoided crossing at $x_2$. At higher interaction strengths $c>c_{\rm chaos}$, the traverse through chaotic intervals results in jagged regions. The sharp chaos-controlled low sweep rate boundary (dashed vertical lines in both panels) becomes more restrictive as $c$ increases, eventually taking over the more moderate $x_2$-controlled power-law degradation.}   \label{f9}
\end{figure}

\begin{figure}[t]
  \centering
   \includegraphics[width=3.5in]{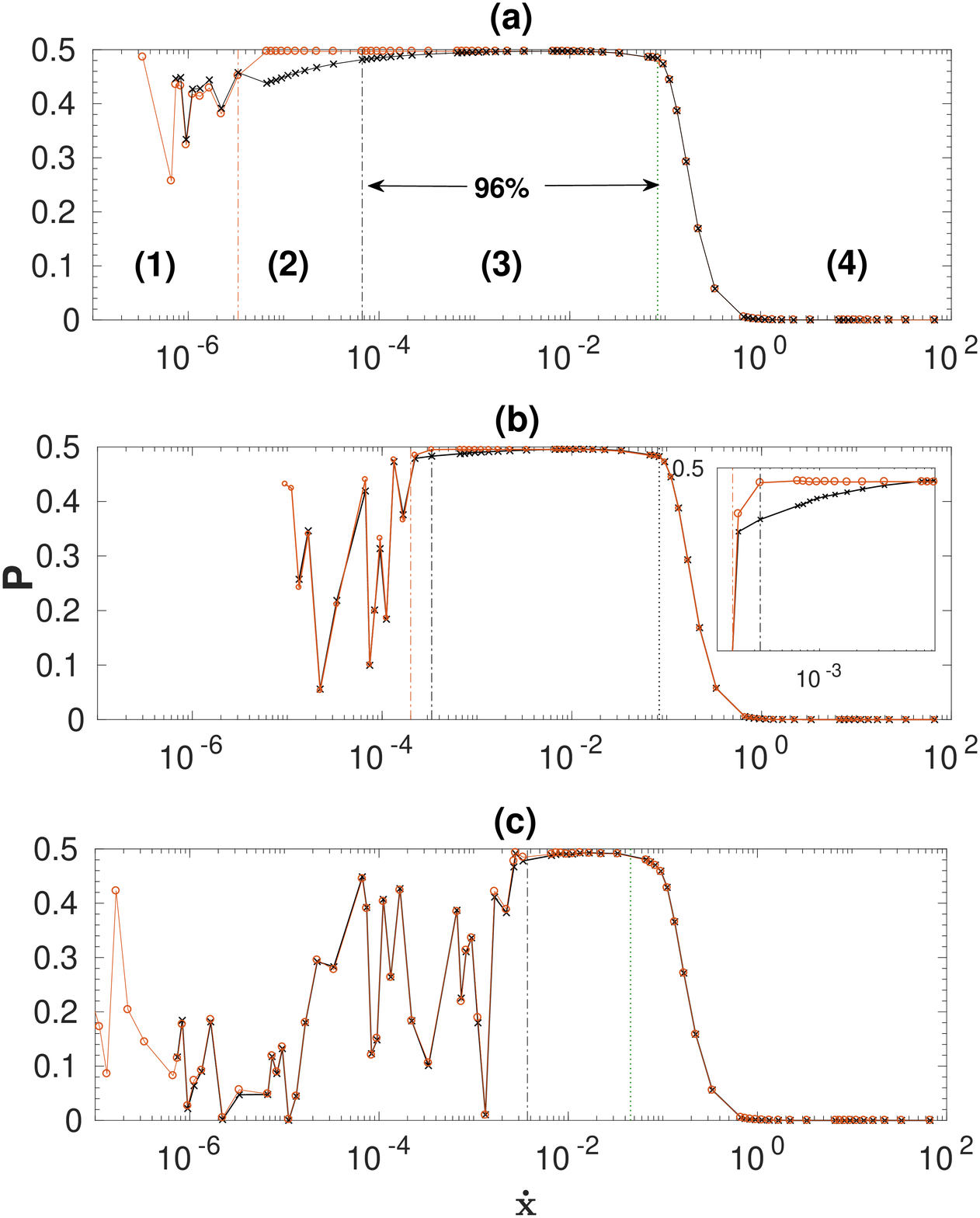}
  \caption{The final conversion yield $P$ is plotted versus the sweep rate ${\dot x}$ for interaction strength values of (a) $c=0.1$, (b) $c=0.2$, (c) $c=0.5$. To isolate the effect of chaos from that of the avoided crossing at $x_2$, we plot both the final efficiency at the end of the full sweep (Black, $\times$) and the efficiency just before $x_2$ (Brown, $\circ$). Vertical lines mark the point where $P$ crosses the 96\% threshold. Dash-dotted lines mark the low rate boundaries, whereas dotted green lines mark the standard high rate boundary. Thus, region (1) in (a) corresponds to chaos-controlled failure, region (2) is the power-law degradation regime controlled by the avoided crossing at $x_2$, region (3) marks the operational high efficiency regime between sweep-rate boundaries, and region (4) corresponds nonadiabatic dynamics where the sweep is no longer slow with respect to the system's characteristic frequencies.}
  \label{f10}
\end{figure}

In Fig.~\ref{f9} we plot the atom-to-molecule conversion efficiency $P=n_g(\infty)$ and the remnant unconverted population $1-2P$ as a function of the sweep rate ${\dot x}$ for various value of the interaction strengths $c$. In the absence of interactions ($c=0$, black solid line) there is no low sweep rate boundary. Interestingly, the remnant population at the slow sweep limit decreases as a {\em power law} $1-2P\propto ({\dot x})^{1/3}$ rather than according to the exponential LZ prescription. This post-LZ behavior is related to the nonlinear atom-pairing term, as detailed in previous studies of interaction-free atom-molecule conversion \cite{Ishkanyan04,Altman05,Pazy05,Tikhonenkov06}. For weak interaction, the avoided crossing appears at $x_2$, but the system remains {\em dynamically} stable throughout the process, i.e. there is no interval of chaotic quasistatic motion. Since the avoided crossing has to be crossed diabatically to achieve high conversion efficiency, we obtain an {\em reciprocal power law} dependence with the remnant population {\em increasing} as $1-2P\propto ({\dot x})^{-1/2}$ with slower sweeps. At stronger interaction, intervals of dynamical instability appear as described in Section~\ref{stability}. The low sweep rate boundary for traversing these chaotic intervals is obtained from the Bogoliubov analysis, as
\begin{equation}
{\dot x} > \frac{\xi}{t_s}~,
\end{equation}
where $\xi$ is the width of the instability interval (the $x$-range where $\mathrm{Im}(\omega)\neq 0$) and $t_s\equiv 1/\max[\mathrm{Im}(\omega)]$ is the characteristic spreading time. These boundaries are marked in Fig.~\ref{f9} and coincide precisely with the breakdown of the power-law dependence and the appearance of jagged disordered features in the efficiency curves. With stronger and broader chaotic intervals at higher values of $c$, the sweep rate required for their successful traverse increases, until it overtakes the weak power law decay due to the $x_2$ avoided crossing and the efficiency becomes chaos-controlled. Due to its narrow width and small gap, the avoided crossing at $x_1$, appearing for $c>0.1$ can only damage the efficiency at much lower values of ${\dot x}$. Thus, like the avoided 'horn' crossings in Refs.~\cite{Dey18,Dey19,Graefe06},  it never controls the low sweep rate boundary. Its effect is significant only for much higher values of $c$ where the adiabatic path from atoms to molecules no longer exists.

The interplay of the two breakdown mechanisms related to the adiabatic following of the avoided crossing at $x_2$ and to the stochastic spreading during the chaotic dynamical instability interval is illustrated in Fig.~\ref{f10}, where we plot efficiency vs. rate curves at stronger ($c>0.1$) interaction. The final efficiency (black, $\times$) involves both effects, but we can separate out the chaos-only degradation, if we stop the process just before $x_2$ (brown, $\circ$).  As also seen in Fig.~\ref{f9}, for $c=0.1$ (Fig.~\ref{f10}a) chaos is only manifest at slower sweep rates (region (1) in Fig.~\ref{f10}a) where there is already significant efficiency deterioration due to the $x_2$ crossing. For this interaction strength, it is still the power-law regime (region (2) in Fig.~\ref{f10}a) that sets the low sweep rate boundary. However, at stronger interaction strengths (Fig.~\ref{f10}b,c) the power-law region shrinks and the final efficiency is purely determined by chaos, as evident from the nearly prefect overlap of the pre-$x_2$ and post-$x_2$ efficiency curves.

\section{conclusion}\label{conclusion}

We have studied the mechanisms degrading the efficiency of atomic-to-molecular BEC conversion via STIRAP in the presence of interparticle interactions. Our findings highlight the significance of dynamical instability and chaos in such processes. In particular, we find that both avoided crossings in the bifurcation diagram and chaotic intervals where the adiabatically followed atom-molecule dark state becomes dynamically unstable without leaving a trace in the bifurcation diagram, can play a role in setting molecular outcomes and introduce {\em low} sweep-rate boundaries for efficient conversion. In the absence of inter-particle interaction, we retrieve the known result of post-Landau-Zener power law decrease of the remnant population as the sweep rate decreases \cite{Ishkanyan04,Altman05,Pazy05,Tikhonenkov06}. In the presence of interactions, in regimes where the conversion efficiency is controlled by an avoided crossing, we find an intriguing inverse power-law dependence of the efficiency on the sweep rate. When chaos limits the conversion, we obtain non-monotonic disordered degradation as the sweep slows down, due to stochastic spreading over the entire energy shell that contains the followed SP. 

\acknowledgements
AV acknowledges support from the Israel Science Foundation (Grant No. 283/18).

\end{document}